\documentclass[submission%
]{dmtcs}

\usepackage[latin1]{inputenc}
\usepackage{subfigure}

\author{Hossein Ghasemalizadeh\addressmark{1}\thanks{Email: \email{ghasemalizadeh@aut.ac.ir}}
  \and Mohammadreza Razzazi\addressmark{1}\addressmark{2}\thanks{razzazi@aut.ac.ir}
  }
\title{Output sensitive algorithm for covering many points}
\address{\addressmark{1}SSRD Lab, Computer Engineering and Information technology department, Amirkabir University of technology, Tehran, Iran.\\
  \addressmark{2}School of Computer Science, Institute for Research in Fundamental Sciences(IPM), P.O. Box:19395-5746, Tehran, Iran.}
\received{1998-10-14}
\revised{\today}
\accepted{tomorrow}
\begin{document}
\maketitle
\begin{abstract}
  A set of points and a positive integer $m$ are given and our goal is to cover the maximum number of these point with $m$ disks.
We devise the first output sensitive algorithm for this problem. We introduce a parameter $\rho$ as the maximum number of points that one disk can cover. In this paper first we solve the problem for $m=2$ in $O({n\rho} + {\rho ^3}\log \rho ))$ time. The previous algorithm for this problem runs in $O({n^3}\log n)$  time. Our algorithm outperforms the previous algorithm because $\rho$   is much smaller than $n$ in many cases. Then we extend the algorithm for any value of $m$ and we solve the problem in $O(m{n\rho} + {(m\rho )^{2m - 1}}\log m\rho )$ time. The previous algorithm for this problem runs in $O({n^{2m - 1}}\log n)$ time.  Our algorithm runs faster than the previous algorithm because $m\rho$ is smaller than $n$ in many cases. Our technique to obtain an output sensitive algorithm is to use a greedy algorithm to confine the areas that we should search to obtain the result. Our technique in this paper may be applicable in other set covering problems that deploy a greedy algorithm, to obtain faster solutions.
\end{abstract}

\section{Introduction}

In the classic covering problem, a set of points are given, and the goal is to place the minimum number of unit disks to cover all the input points. In the maximum coverage problem, the number of disks to be used, $m$, is given and the goal is to cover the maximum number of points with $m$ disks. We call this problem $MostPoints(P,m)$. This problem was introduced by Drezner \cite{1} for $m=1$ and it was solved in $O({n^2}\log n)$ time. To solve the problem, he replaced every point with a disk centered at that point and he obtained the maximum depth in the arrangement of disks. Later, Chazelle and Lee \cite{2}  solved the problem in $O(n^2)$ time. This problem belongs to the 3-SUM hard complexity class \cite{3}, which means that,  the $O(n^2)$ running time algorithm is the best one for this problem. However, there are some algorithms which approximately solve this problem in less than $O(n^2)$ time. These algorithms approximate the disk radius and the number of covered points. Figueiredo and Fonesca \cite{5} gave an algorithm to cover the maximum number of points that a unit disk in ${\Re^d}$   can cover, with a  $(1 + \varepsilon )$-radius disk  in $O(n/{\varepsilon ^{d - 1}})$   time. In  \cite{4}  Aronov and Har-peled gave a  $(1 - \varepsilon )$-approximation algorithm in the number of covered points which runs in $O(n{\varepsilon ^{ - 2}}\log n)$ time.
$MostPoints(P,m)$, for $ m > 1 $, is NP-Hard unless we consider $m$ as a constant. A trivial greedy algorithm is a  $(1 - \frac{1}{e})$-approximation algorithm for it \cite{6}. The greedy algorithm first finds a disk which covers the maximum number of points in $O(n^2)$  time. To pick the next disks, it removes the points located in the first disk, and finds the disk which covers the maximum total weight of points as the second disk. It repeats this process until $m$ disks are picked up. This yields a  $(1 - \frac{1}{e})$-approximation algorithm which runs in $O(m{n^2})$  time. The first $(1 - \varepsilon )$-approximation algorithm  for this problem was given in \cite{7} which runs in $O(n\log n + n{\varepsilon ^{ - 6m + 6}}\log (\frac{1}{\varepsilon }))$ time. We presented a polynomial time approximation schema for this problem in our previous work \cite{8}, which runs in $O((1 + \varepsilon )mn + {\varepsilon ^{ - 1}}{n^{4\sqrt 2 {\varepsilon ^{ - 1}} + 2}})$ time.

The trivial method to obtain the optimal result of the $MostPoints(P,m)$ is to consider all subsets of size $m$ of $n^2$ possible disks and finds the subset which covers the most number of points. This method takes $O(n^{2m})$ time. In \cite{7} de Berg et. al gave an   algorithm for this problem. They first solved the problem for $m=2$ in $O({n^3}\log n)$   time. To solve the problem for $m>2$, they fixed every subset of $m-2$ disks, and they found the best 2 disks after removing the points contained in the $m-2$ disks. This takes $O({n^{2m - 1}}\log n)$  time. \\
In section 2, we present an output sensitive algorithm to obtain the optimal result of  $MostPoints(P,2)$ which runs in $O({n\rho} + {\rho ^3}\log \rho )$ time where $\rho$  is the maximum number of points that one disk can cover. In Section 3 we extends the algorithm for $MostPoints(P,m)$ . In section 4 we compare the implementation results of our algorithm and the algorithm of \cite{7} and in section 5 we conclude the paper.  Our innovation in this algorithm is to use the greedy algorithm to find the regions in which the $m$ resultant disks may reside, and searching in those regions only. In the rest of the paper when we use disks we mean unit disks.

\section{Output sensitive algorithm for MostPoints(P,2)}
 In this section we describe our algorithm for $MostPoints(P,2)$. Let $g_1 =MostPoints(P,1)$ be the disk that covers the maximum number points from the point set $P$. Let $Points(g_1)$ refers to the points of $P$ which are located inside $g_1$. Define  $g_2=MostPoints(P - Points(g_1),1)$ as the disk that covers the maximum number of points after removing the points located in $g_1$. $g_1$ and $g_2$ are the result of the greedy algorithm for $MostPoints(P,2)$. An example is given in Figure 1. In this example, the optimal solution covers 18 points whereas the greedy algorithm covers 15 points. As illustrated in this example, the two disks of the optimal solution have common points with $g_1$. Intuitively, either the disks of the optimal solution have common points with $g_1$ or, the greedy algorithm obtains the optimal solution. Lemma 1 proves this claim.
\begin{figure}[htbp]
\begin{center}
   \includegraphics[width=.5\columnwidth]{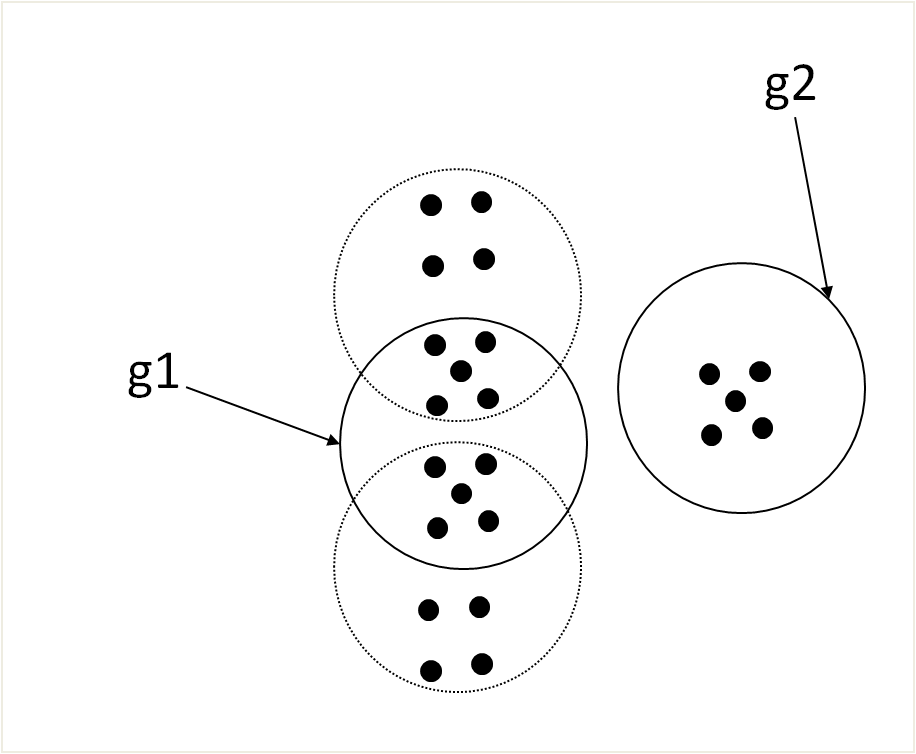}
  \caption{This figure shows the greedy solution and the optimal solution for $MostPoints(P,2)$, in a sample point set. In this example, the greedy algorithm returns $g_1$ and $g_2$ which cover 15 points together, whereas the disks in the optimal solution, the  two disks specified with the dash lines, cover 18 points together. }
  \label{fig:fig1}
  \end{center}
\end{figure}
By the best two disks we mean the optimal solution of $MostPoints(P,2)$. Let $D$ be a set of disks. The function $Cover(D)$ denotes the number of points covered by the disks in $D$.

\textbf{Lemma 1}: Let $g_1$ be one of the disks that covers the maximum number of points and $g_2$ be one of the disks that covers the maximum number of points after removing the points located in $g_1$. Let    $o_1$ and $o_2$ be the two disks that together cover the maximum number of points among any combination of two disks: Either both $o_1$ and $o_2$ have common points with $g_1$, or $g_1$ and $g_2$ cover the maximum number of points.

\begin{proof}
 Suppose that at least one of $o_1$ and $o_2$, say $o_2$, does not have any common point with $g_1$. Then we have $Cover(\{g_1, g_2\}) \geq Cover(\{g_1 , o_2\})$, because $g_2$ covers the maximum number of points not covered by $g_1$. We also have $Cover(\{g_1,o_2\}) \geq Cover(\{o_1 , o_2\})$ because $g_1$ covers at least the same weight of points as $o_1$ (since it is a disk that covers the maximum weight of points) and $g_1$ does not have any point in common with $o_2$. Thus $Cover(\{g_1 , g_2\}) \geq Cover(\{o_1, o_2\})$, which implies that the total number of the points covered by $g_1$ and $g_2$ is maximal.
\end{proof}

Based on Lemma 1, we should look for the best two disks in the disks which have common points with $g_1$. These disks are located around $g_1$ in a circle having the same center as $g_1$ and radius $3$, where the radius of $g_1$ is 1. We call this region $Ng_1$. Figure 2  shows $Ng_1$. Lemma 2 bounds the number of points in $Ng_1$.
\begin{figure}
\begin{center}
   \includegraphics[width=.5\columnwidth]{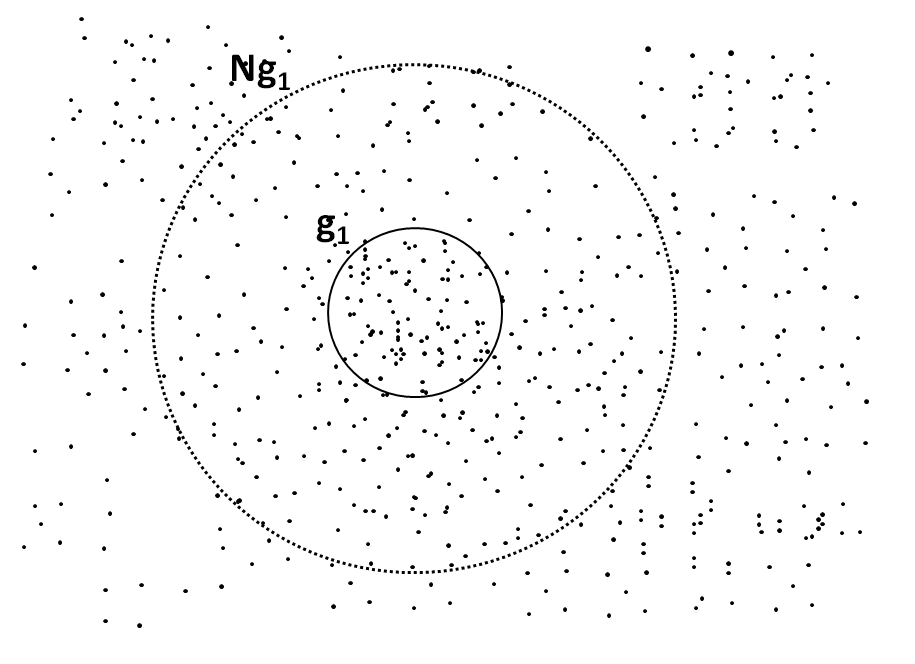}   
  \caption{The disks which have common points with $g_1$ are located in the dashed region around $g_1$. We call this region $Ng_1$}
  \label{fig:fig2}
  \end{center}
\end{figure}

\begin{figure}
 \begin{center}
   \includegraphics[width=.5\columnwidth]{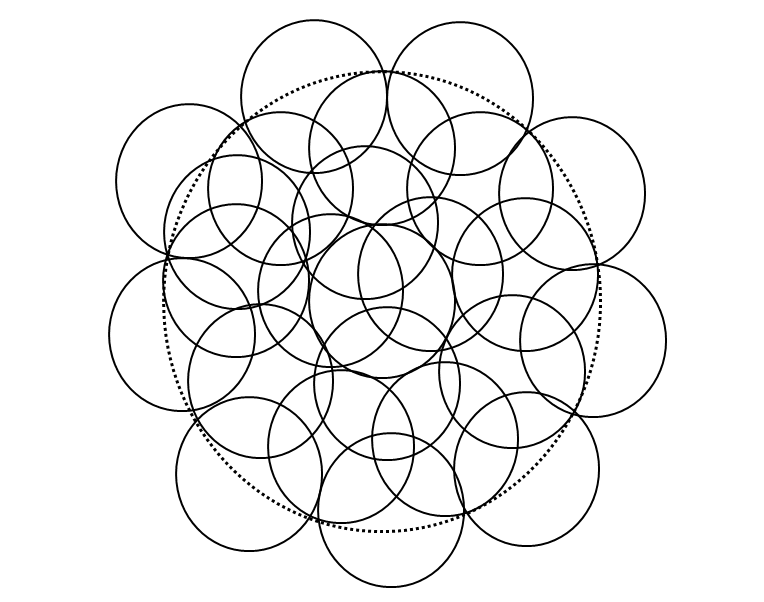}   
  \caption{We can pack the dashed region with at most 21 unit disks}
  \label{fig:fig3}
  \end{center}
\end{figure}

\label{lem:Lemma 2}
\textbf{Lemma 2}: Let $\rho$ be the maximum number of points that can be covered by a disk. The number of points in $Ng_1$ is $O(\rho)$.

\begin{proof}
 We can pack $Ng_1$ with at most 21 unit disks, which is shown in Figure ~\ref{fig:fig3}. Every one of these 21 disks covers equal or less than $ \rho$ points. So the number of points in $Ng_1$ is at most $21\rho $ points, which is $O(\rho)$ points.
\end{proof}
\\We can also devise an output sensitive algorithm  for $MostPoints(P,1)$ that solves the problem in $O(n\rho)$ time. The algorithm is as follows. Cover the plane using four shifted grids $G_1,...,G_4$. Each grid covers the whole plane, but the grids are shifted relative to each other, 2 units from each side. The cells of the grids have size 4x4. For any unit disk $d$ there is a grid $G_i$ such that $d$ is fully contained in a single cell of $G_i$. Now, for each non-empty grid cell $C$ of every grid, compute the optimal disk on the set of points lying inside $C$.  $C$ contains at most  $c\rho$ points, for a constant value of $c$ and the time spent for cell $C$ is $O(|C|^2)$ where $|C|$ denotes the number of points in the cell. The total number of points over all cells $C$ is $4n$, since each point is contained in exactly one cell per grid. To obtain the running time of the algorithm, we count the cells of the grids $C_1,...,C_l$.  The total running time of the algorithm is:\\
 $\sum\limits_{i = 1}^l {|{C_i}{|^2}}  \le \sum\limits_{i = 1}^l {|{C_i}| \times c\rho  \le c\rho  \times 4n = O(\rho n)} $
\\
Using  Lemma 1 and Lemma 2, the best two disks can be $g_1$ and $g_2$, or the two disks around $g_1$ that cover the maximum number of points. To find the best two disks in $P$, first we find $g_1$ and $g_2$ in $O(n\rho)$  time. Then, we determine the points in $Ng_1$, and we run the algorithm of \cite{7} for the points located in $Ng_1$. This algorithm obtains the best two disks in $Ng_1$, in $O(\rho^3\log\rho)$   time. Finally, we compare this result, the best two disks in $Ng_1$, with $g_1$ and $g_2$ and consider the one that covers more weight as the optimal solution for $MostPoints(P,2)$. Since $\rho$   can be much smaller than $n$ in practice, our algorithm which runs in $O({n\rho} + {\rho ^3}\log \rho )$ time, is an improvement over the algorithm of \cite{7} that requires $O({n^3}\log n)$  time.

\section{The algorithm for MostPoints(P,m)}
The algorithm can be extended to cover the maximum number of points with $m$ disks. From now on, by the best $j$ disks, we mean a set of $j$ disks that cover the maximum total weight of points. The proposed algorithm is an iterative algorithm. At the first iteration, the algorithm finds two disks, which cover the maximum number of points as explained before. To find the best 3 disks, the algorithm uses the following fact, which will be proven in Lemma 3: the best 3 disks are among the disks that have some common points with the best 2 disks; Otherwise, the best 2 disks plus one disk that covers the maximum number of points, after removing the points located in the best two disks, are the best 3 disks. The algorithm repeats this process until the best $m$ disks are found. Now we explain the algorithm formally. Remember that the function $MostPoints(Q,j)$ returns  a set of $j$ disks that cover the maximum number of points in the point set $Q$. Let $D$ be a set of unit disks. Define $Points(D) = \{ p|p \in (P \cap D)\} $  as the subset of $P$ that is located in the disks of $D$. Define $Neighbor(D)$ as the points in the circles with the same centers as the disks in $D$ and radius $3$. Let $OPT_i$ denotes the best $i$ disks.
The outline of the algorithm is as follows:

\textit{
 \\
The exact output sensitive algorithm for the most points covering problem
 \\Inputs: A set P of n points in the plane, and a positive integer m.
 \\Outputs: A set of m disks which cover the maximum number of points from P.
 \\Definitions: The function MostPoints(Q,j), Points(D), and Neighbor(D) as defined in the text.}
 \\\\
 1 \hspace{2 mm} begin\\
 2 \hspace{4 mm} $OPT_0= \emptyset $   \\
 3 \hspace{4 mm}$g_1= MostPoints(P,1) $\\
 4 \hspace{4 mm}$OPT_1 = g_1 $\\
 5 \hspace{4 mm}For $i=2$ to $m$  \\
 6 \hspace{8 mm} $rp= P-Points(OPT_{i-1})$\\
 7 \hspace{8 mm}$g_i= MostPoints(rp,1)$\\
 8 \hspace{8 mm}$O_i= MostPoints(Neighbor(OPT_{i-1}), i)$ \\
 9 \hspace{8 mm}If ( $Cover(OPT_{i-1}  \cup g_i) > Cover(O_i)$ ) \\
 10\hspace{12 mm} $OPT_i= OPT_{i-1} \cup g_i$ \\
 11\hspace{7 mm} else  \\
 12\hspace{12 mm}$OPT_i = O_i$ \\\\
 13\hspace{3 mm} return $OPT_m$ \\
 14\hspace{2 mm}end \\

In Lemma 3 we prove the correctness of the above algorithm.  Before proving Lemma 3 we have the following definitions.\\
\textbf{Definition 1}:  Let $D$ and $E$ be two sets of  disks.  Define $ExculsiveCover(D,E)$ as the function which returns the number of points covered by $D$ after removing the points covered by $E$. By the definition of $ExculsiveCover(D,E)$ and $Cover(D)$ functions we have the following facts:
\[ExculsiveCover\left( {D,E} \right){\rm{ }} \le Cover(D)\]
\[Cover(D \cup E) = Cover(D) + ExclusiveCover(E,D)\]
\[Cover(D \cup E) \le Cover(D) + Cover(E)\]

\textbf{Lemma 3}: The exact output sensitive algorithm for the most points covering problem obtains $m$ disks that cover the maximum number of points.

\begin{proof}
We prove the correctness of the algorithm by induction on the number of iterations.  Assume that at the end of iteration $i$ we have correctly computed $OPT_i$. Consider $OPT_{i+1}$, which consists of $i+1$ disks. These disks may have common points with $Points(OPT_i)$ or not. We consider both cases. In case 1 we consider that all $i+1$ disks of $OPT_{i+1}$ have common points with $Points(OPT_i)$, and in case 2 we consider that some of these $i$ disks do not have any common point with $Points(OPT_i)$.\\\\
 Case1: All $i+1$ disks of $OPT_{i+1}$ have common points with $Points(OPT_i)$:\\\\
 All disks that have common points with $Points(OPT_i)$, have some points in $Neighbor(OPT_i)$. The algorithm, in line 8, obtains the best $i+1$ disks in the points of $Neighbor(OPT_{i})$. So in this case the algorithm exactly computes $OPT_{i+1}$.
\\\\Case 2:  Some of the disks in $OPT_{i+1}$ do not have any common point with $Points(OPT_i)$:\\\\
 Call one of these disks $F$. We can partition $OPT_{i+1}$ to $\{ OPT_{i + 1} - F\} $  and $F$. Based on the Definition 1, we have:\\\\
$\begin{array}{l}
OPT_{i + 1} = \{ OPT_{i + 1} - F\}  \cup \{ F\}  \\
Cover(OPT_{i + 1})  =  ExclusiveCover(OPT_{i + 1},F) \\
+ Cover(F)  (1)
\end{array}$        \\

 $\{ OP{T_{i + 1}} - F\} $ is a set of $i$ disks. As $OPT_i$ is a set of $i$ disks which cover the maximum number of points, we have:
\\
  $Cover(OP{T_{i + 1}} - F) \le Cover(OP{T_i})$       (2)\\\\
Furthermore, $g_{i+1}$ is the disk that covers the maximum number of points after removing the points in $Points(OPT_i)$. Therefore, $g_{i+1}$ covers the maximum number of points among the disks that have not common points with $OPT_i$. Thus, we have:
 \\ $Cover(F)\; \le \;Cover({g_{i + 1}})$    (3)\\\\
From (2) and (3) we have:
\\\\$Cover(\{ OP{T_{i + 1}} - F\} ) + Cover(F)\; \le \;Cover(OP{T_i}) + Cover({g_{i + 1}})$ (4)
\\\\From (1) and (4) we have:
\\\\ $Cover(OP{T_{i + 1}}) \le \;Cover(OP{T_{i - 1}}) + Cover({g_{i + 1}})$\\

Thus in case 2, $OPT_i$ and $g_{i+1}$ cover the maximum number of points.\\
The algorithm compares the results of case 1 with the result of case 2, and considers the best one as $OPT_{i+1}$. So, the algorithm correctly computes $OPT_{i+1}$. The base case, $i=1$, is correct and can be satisfied using the algorithm devised for $MostPoints(P,1)$.

\end{proof}

In the following lemma, we obtain the running time of the algorithm.

\textbf{Lemma 4}: The exact output sensitive algorithm for the most points covering problem runs in $O(m{n\rho} + {(m\rho )^{2m - 1}}\log m\rho )$ time, where $\rho$   is the maximum number of points that one disk can cover in the points set $P$.

\begin{proof}

The algorithm runs in $m$ iterations. At the start of the iteration $i$, we have found the best $i-1$ disks, and we are getting to find the best $i$ disks. To find the best $i$ disks, we obtain $g_i$ in line 7, which takes $O(n\rho)$   time. We also look for the $i$ disks covering the maximum number of points in $Neighbor(OPT_{i-1})$, in line 8. $Neighbor(OPT_{i-1})$ consists of $i-1$ disks, and based on Lemma 2, the neighborhood of each disk can be packed with at most 21 disks. So the maximum number of points in $Neighbor(OPT_{i-1})$ is $21\rho(i - 1)$,  which is $O(i\rho)$ points. We search for the $i$ disks that cover the maximum number of points in  $O(i \rho )$  points. The algorithm of \cite{7} is applied for these    points, which takes $O({(i \rho )^{2i - 1}}\log (i\rho ))$ time. The algorithm runs in $m$ iterations so, the running time is bounded by $O(n\rho) + \sum_{i=2}^{m} ( O(n\rho) + (i\rho)^{2i-1} \log (m\rho) )$ which is $O( mn\rho + (m\rho)^{2m-1} \log(m\rho))$.\\
\end{proof}

\section{ Implementation results}
We have implemented  our algorithm and the algorithm of \cite{7} for $MostPoints(P,2)$. In each algorithm,  we observed the number of  pairs of disks that each algorithms consider to obtain the optimal solution. We distributed different number of points randomly in squares with different side length and we run both algorithms on the points. Table 1 shows the result of the comparison. The results show that our algorithm computes significantly less number of pairs than the algorithm of \cite{7} in most cases. When $\frac{\rho }{n}$   becomes close to 1, the number of pair of disks that are considered in both algorithms become close to each other.
\section{Extending the algorithm for other shapes and weighted point sets}
This algorithm can be applied for other types of maximum coverage problem such as covering point sets with other shapes and maximum coverage problem in sets. The algorithm uses the parameter $\rho$ which is the cardinality of the largest subset. Lemma 1 and 3 can be applied for general set systems. So we should search in the subsets that have common members with the largest subsets. To bound the running time of the algorithm there should be a constant number of subsets that have common points with the largest subsets.

\section{Conclusion and future works}
In this paper we presented an exact output sensitive algorithm for covering many points problem. In the algorithm, we confined the region in which the resultant disks may reside using the greedy algorithm, and we searched in that regions only. This led to an output sensitive algorithm for this problem which runs in $O(m{n\rho} + {(m\rho )^{2m - 1}}\log m\rho )$  time, where $\rho$  is the maximum number of points that one disk can cover.
Our algorithm can be improved by improving the running time of the algorithm in \cite{7}. If  better algorithms for covering most points with two disks are suggested, they can improve the running time of our algorithm and the algorithm of \cite{7}. Furthermore, considering $\rho$  as a  parameter, and using our technique, it may be possible to improve the analysis of other covering algorithms.
\section{Acknowledgement}
We would like to acknowledge anonymous referees for their helpful comments and the improvements they suggested for the paper.

\begin{table}
\caption{ Number of pairs of disks processed in our algorithm an algorithm of [7] }
\centering          
\begin{tabular}{ |l|l|l|l|l|}    
\hline\hline                        
Num of points & square side & $\rho$ & Num of pairs in algorithm of \cite{7}
 & Num of pairs  in our algorithm \\ [.5ex]  
\hline
1000& 200 &19	&7007&201 \\
1000&100	&44	&22878	&1131\\
1000&50 &	125&	45414&	4259\\
500&50	&105&	26041&	6402\\
500&100	&34&	8605&	252\\

\hline
\end{tabular}
\label{table: Table 1}
\end{table}


\small
\bibliographystyle{abbrv}

\end{document}